# Where is Synergy Indicated in the Norwegian Innovation System?

# Triple-Helix Relations among Technology, Organization, and Geography

Øivind Strand [i]

Aalesund University College, Department of International Marketing, PO Box 1517, 6025 Aalesund, Norway; +47 70 16 12 00; ost@hials.no

Loet Leydesdorff

University of Amsterdam, Amsterdam School of Communication Research (ASCoR), Kloveniersburgwal 48, 1012 CX Amsterdam, The Netherlands; loet@leydesdorff.net

**ABSTRACT**
Using information theory and data for all (0.5 million) Norwegian firms, the national and regional innovation systems are decomposed into three subdynamics: (*i*) economic wealth generation, (*ii*) technological novelty production, and (*iii*) government interventions and administrative control. The mutual information in three dimensions can then be used as an indicator of potential synergy, that is, reduction of uncertainty. We aggregate the data at the NUTS3 level for 19 counties, the NUTS2 level for seven regions, and the single NUTS1 level for the nation. Measured as in-between group reduction of uncertainty, 11.7 % of the synergy was found at the regional level, whereas only another 2.7% was added by aggregation at the national level. Using this triple-helix indicator, the counties along the west coast are indicated as more knowledge-based than the metropolitan area of Oslo or the geographical environment of the Technical University in Trondheim. Foreign direct investment seems to have larger knowledge spill-overs in Norway (oil, gas, offshore, chemistry, and marine) than the institutional knowledge infrastructure in established universities. The northern part of the country, which receives large government subsidies, shows a deviant pattern.

---

[i] Corresponding author

# 1.    Introduction

Innovation takes place in a landscape of interactions, collaboration, and knowledge exchanges among firms, academic institutions, and various government agencies [1]. Firms and institutional agents cooperate and participate in networks at various geographical scales; locally, regionally, nationally, and internationally [2]. Whether and how government interventions, or the presence of academia, matter for regional innovation is an issue of political significance in many countries because innovation in the regions is considered to be a condition for increasing prosperity [3-5]. Accordingly, national and regional governments in several countries have developed programs and centres for enhancing innovation in the regions [6, 7]. A number of factors are important in this context: the industry structure [8], the role of the universities [9, 10], the role of knowledge networks [2, 11], proximity and localization [12-15], and organization and culture [16, 17].

Leydesdorff and Meyer [18] raised the question of how to measure whether a knowledge base in the economy is developed more at the regional than the national level (or *vice versa*). Can something as elusive as the knowledge base of an economy be measured in terms of the interactions in a Triple Helix between economic development, organized knowledge production, and political control? The purpose of this paper is to estimate the characteristics of such Triple-Helix dynamics in the Norwegian innovation system. Combining the use of information theory and the Triple-Helix model of university-industry-government relations, we propose a tool for measuring the extent to which innovations have become systemic.

Canter *et al.* [2], for example, used patent data from firms in three industrial regions to characterize the knowledge base of the regions. Our approach provides an empirical alternative to the *a priori* assumption that such systems would exist geographically either at the national or regional levels. We use an information-theoretical method on a complete set of micro-level data for all—that is, almost half a million—Norwegian firms registered during the last quarter of 2008. Each of these firms is attributed a municipality code (as a proxy for geography), a sector code (proxy for technology), and a size code for firms (proxy for organization).

The study leans on three previous papers using a similar method, but containing data from the Netherlands [19], Germany [20], and Hungary [21]. These studies have similarities in their methodological approach, but were different in several ways. The Hungarian study focused on firms from high-tech industries and knowledge-intensive services. The German study did not contain data about self-employed firms. The study of the Netherlands used postal codes instead of municipalities as the geographical proxy. Furthermore, the geography and the industry patterns in Norway are different from the other countries studied. The state can be expected to play a more active role in Norway than in the other countries for which similar studies were performed [1: p. 111].

This study broadens the picture from previous studies by including two new elements in the analysis. First, by including the geographical distribution of foreign factors [22, 23], such as foreign direct investment and export incomes (at the county level). Second, by discussing the distribution of research funding among Norwegian counties. Following Leydesdorff *et al*. [19], we first combine the theoretical perspective of regional economics [24] with the Triple-Helix model [1]. Three dimensions are thus distinguished: technology, geography, and organization. These dimensions cannot be reduced to one another, but interactions among them in networks of university-industry-government relations can be expected. The synergy in these interactions can be measured in the Norwegian innovation system and can also be decomposed at different levels of scale [25].

The mutual information among the three dimensions (geography, technology, and organization) can be negative and can then be interpreted as an indicator of reduction of uncertainty or synergy. Lengyel & Leydesdorff [21] specified the synergetic functions as 'knowledge exploration' (between technology and geography), 'knowledge exploitation' (between technology and organization) and 'organization control (between organization and geography). Spurious correlations among these interacting subdynamics of a knowledge-based system may reduce the uncertainty that prevails, and this reduction can be measured using the mutual information in three dimensions. Yeung [26] specified the resulting indicator as a signed information measure. A signed measure can no longer be considered as a Shannon entropy [27].

When this signed information measure is negative, the synergy among the functions reduces uncertainty that prevails at the systems level. The synergy is an attribute to the configuration, and not of the composing subdynamics. It emerges as a virtual knowledge base that feeds back on the composing subdynamics. However, information theory allows for the precise decomposition into components of this knowledge base in terms of bits of information [25].

We study the measure at four geographical levels: the national system (NUTS1),[1] seven regions (NUTS2), 19 counties (NUTS3), and 430 municipalities (NUTS5). The results enable us to specify where synergy is highest and whether the respective innovation systems have more regional or national characteristics.

Etzkowitz and Leydesdorff [1; p. 111] used Norway as an example for the Triple-Helix I model, where the strong state governs academia and industry. Onsager *et al.* [28] reported that the largest city regions in Norway seem to have limited capacity to utilize their resource advantages and potential synergy. Herstad *et al.* [29] concluded that firms in the capital region (Oslo) are less engaged in innovative collaboration than firms in the rest of the country, whereas Isaksen and Wiig Aslesen [30] argued that the knowledge organizations in Oslo do not (yet) function as hubs in a wider innovation system.

The relations between innovation, policy, and inter-firm linkages in Norway were also discussed by Nooteboom [31]. He concluded that central government should limit itself to facilitation in the formation of enterprise clusters. An OECD report [32], analyzing the roles of knowledge institutions in the Trondheim region, concluded that in spite of being Scandinavia's largest independent research institution and technical university, there is a need to 'broaden the innovation dynamics' and increasing the absorptive capacity within this region. The existence of fragmentation [28] and 'parallel worlds' [32] within the Norwegian innovation system, can be considered as indications of redundancy rather than synergy.

---

[1] *NUTS* is an abbreviation for *Nomenclature of territorial units for statistics*.

In this study, we address these Triple-Helix issues empirically by using data and information theory. We focus on the geographical decomposition of the configurations. The main research question is to find and explain geographical areas where synergy among the knowledge-based innovation functions is higher than in other areas. From a methodological perspective, it is interesting to study first the complete populations of firms, that is, without focusing on sectors or geographical areas which are *a priori* defined as relevant systems of innovation. The finely grained geographical mesh of the Norwegian firms allows us to estimate at which geographical levels synergies occur. Additionally, we relate our results to the geographical distribution of government spending on R&D and foreign factors in areas of high or low synergy. Finally, we also reflect and elaborate on some counter-intuitive results.

## 2. Theoretical perspectives

Storper [24] defined a territorial economy as a 'holy trinity' of relational assets. In figure 2.3 on page 49 of his study (see Figure 1 below), the economy is considered as a set of intertwined, partially overlapping domains of action. The building blocks of this 'holy trinity' are technology, organizations, and territory (geography). There are three bilaterally overlapping domains between the three spheres and one trilateral. A domain where technology and organizations overlap is then characterized as the 'world of production' with a specific 'system of innovation.' The domain where organizations and territory overlap is denoted by Storper as the 'regional world of production.' The domain where technology and territory overlap is called the 'regional world of innovation.' The trilateral domain is a combination where the three bilateral domains overlap.

Analogously, the Triple-Helix model of university-industry-government relations shows bi- and trilateral characteristics of overlap [1, 33]. These two corresponding models can both be misinterpreted as static. However, Figure 1 provides a picture 'frozen in time' whereas the subsystems evolve over time. The time axis is perpendicular to the paper-plane, forming helices in a complex and nonlinear way. Each of the subdynamics of these models interacts with the two others and with itself [34], and they can represent sub-dynamics at various scales. As noted by Leydesdorff *et al*. [19, 35] and with reference to Storpers' original figure, a gap in the overlap between the three circles can also be understood as a representation of negative information—that is, reduction of uncertainty or, in other words, unintended synergy among the three heterogeneous fluxes.

Figure 1: Overlap between the three institutional spheres, indicating positive and negative information in the trilateral overlap.

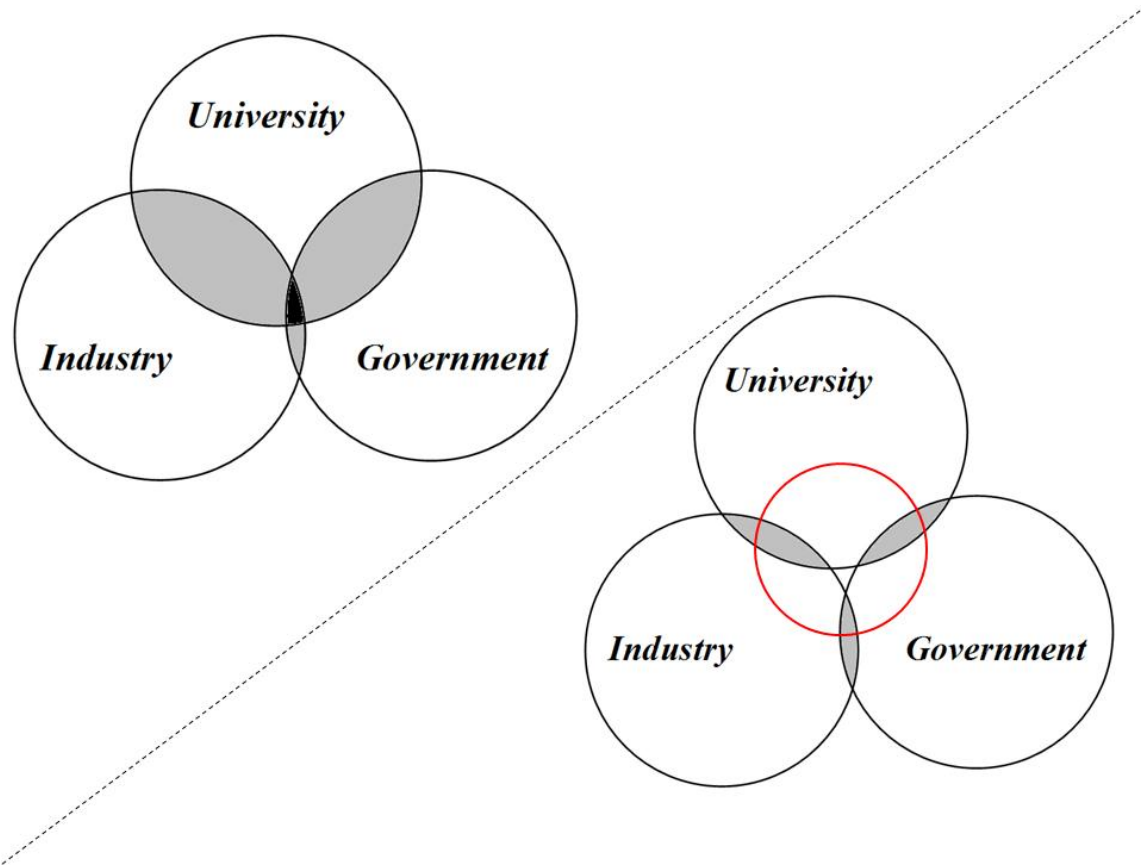

The Triple-Helix model first provides a neo-institutional model for studying the network arrangement among the three different types of agents in university-industry-government relations [1]. Adding the functions to the neo-institutional model, the latter can further be developed into a neo-evolutionary model [33] with emphasis on the relations among the (latent) functions that operate in and on the networks. Each knowledge function is represented as a bilateral interaction term; for example, "knowledge exploitation" between the dimensions technology and organization. Synergy among the three knowledge functions can then be considered as an indicator of the quality of an innovation system.

The use of these uncertainty measures [36, 37] has gained much attention in various areas of study in recent years. Guo [38], for example, used entropy methods for geographical data mining; Boschma & Iammarino [39] used similar methods in a study about trade linkages and regional growth in Italy. Chanda *et al.,* [40, 41] used these techniques on bio-informatics for the visualization of Gene-Environment Interactions [42]. According to these authors, the association information between attributes of data sets provides insight into the underlying structures in the data [40].

Association information can broadly be categorized into correlation information and interaction information. The correlation information among the attributes of a data set can be interpreted as the total amount of information shared between the attributes. The interaction information can be interpreted as multivariate dependencies among the attributes. A spurious correlation in a third attribute can reduce the uncertainty between the other two. Compared with correlation, this 'mutual information' can be considered as a more parsimonious measure for the association at the systems level. The concept of mutual information among three dimensions was first introduced by McGill [43] as a generalization of Shannon's mutual information [44] to more than two dimensions. The measure is similar to the analysis of variance, but uncertainty analysis remains more abstract and does not require assumptions about the metric properties of the variables [45]. Han [46] further developed the measure, and positive and negative interactions were also discussed by Tsjishita [47] and Yeung [26; p.59].

*2.1 Characteristics of the data about Norway*

The context of this study is the Norwegian economy, which features a combination of free-market activities and government interventions. The public sector is, as in the other Scandinavian countries, relatively large in comparison to other European nations. Furthermore, Norway is among the few European countries that are not members of the EU. The country is richly endowed with natural resources—petroleum, hydropower, fish, forests, and minerals—and is highly dependent on the petroleum sector [48]. As in other advanced industrial nations, Norway is engaged in an ongoing transformation from resource-based to knowledge-based industries [49].

Norway is one of Europe's most mountainous countries and has a rugged coastline with almost 50,000 islands. These geographical conditions are the main reason for the large number of small municipalities. There are only five urban settlements with a population of more than 100,000 inhabitants: the capital Oslo, Bergen in Hordaland, the Stavanger/Sandnes area in Rogaland, Trondheim in Sør-Trøndelag, and the Fredrikstad/Sarpsborg area in Østfold [50]. In Oslo, almost 100% of the population lives in urban settlements, whereas these figures are 60% in Hedmark, Oppland, Sogn og Fjordane, and Nord-Trøndelag. The country has a population of 4.9 million and the population density is 15.8 inhabitants/km$^2$ [51]. This is amongst the lowest in Europe. For example, the population densities for the other countries where such a Triple-Helix analysis was performed, were: Hungary (107.9), Germany (229.9), and the Netherlands (487.2).

Norway is organized at three levels of government: the central government (NUTS1), 19 counties (at the NUTS3 level) and 430 municipalities (at the NUTS5 level). In addition to these administrative levels, the country can for statistical reasons be divided into 90 economic regions at the NUTS4 level and seven regions at the NUTS2 level. The firm data in this study are specified at the (lowest) municipality level (NUTS5). Note that Oslo is considered as both a county and a municipality. We do not use the economic regions at the NUTS4 level because this would mean that data from these regions inside Oslo could then not be extracted. We perform our analyses at the county level (NUTS3) and at the level of regions (NUTS2).

A further study of qualitative characteristics of various regional innovation systems in Norway can be found in [52, 53]. Onsager *et al.* [28] focused on the city regions in Norway, and Asheim & Conen [54] focused on the knowledge base of regional innovation systems in the Nordic countries. Narula [55] investigated innovation systems and 'inertia' in R&D locations in Norway. Isaksen and Onsager [56] analyzed the knowledge-intensive industry in Norway. Isaksen [7] investigated the innovation dynamics of six regional clusters in Norway. He identified a micro system cluster in Vestfold, a systems engineering cluster in Buskerud (Kongsberg), a light-metal cluster in Oppland (Raufoss), a subsea cluster in Hordaland (Bergen), a maritime cluster in Møre og Romsdal, and an instrumentation cluster in Sør-Trøndelag (Trondheim).

## 2.2 *Characteristics of the knowledge infrastructure in Norway*

The knowledge infrastructure of Norway is young, distributed, and rapidly changing. Currently, the country has eight universities located in Oslo (founded in 1811), Bergen in Hordaland (1946), Stavanger in Rogaland (2005), Agder (2007), Tromsø in Troms (1968) and Bodø in Nordland (2011). The only technical university is located in Trondheim in Sør-Trøndelag (founded in 1910). The University for the Life Sciences is located in Ås in Akershus (founded in 1859). The Norwegian School of Economics, located in Bergen, was founded in 1936.

A number of 26 small, state-owned, university colleges are located in almost every county. There is an ongoing process of fusion between these colleges to form universities or larger units. The economic and political freedom to self-organize the economy, the rapid transition of industries and knowledge infrastructures, and the relatively high level of governmental interventions make Norway an interesting case for a Triple-Helix analysis [1].

## 3. Methods and data

*3.1 Data*

The data consist of information about 481,819 firms, provided by Statistics Norway. The figures were collected for the fourth quarter of 2008 and were harvested from the web site of Statistics Norway [57]. These data cover the complete population of Norwegian firms. All records contain the three variables which we can use as proxies for the dimensions of geography, technology, and organization. Geography is indicated by a four-digit code for municipalities; these data can be aggregated under a two-digit county code and a one-digit regional code. The municipality is the lowest level of analysis (NUTS5) and the lowest level of administration in Norway. The counties are used as the second level of administration at the NUTS3 level. The regional level (NUTS2) is also included in our analysis, even though it does not represent a separate level of administration in Norway.

Table 1: Geographical subdivision of Norway.

| NUTS2 Code | Regions | NUTS3 code | County | NUTS3: number of firms | NUTS5: number of municipalities |
|---|---|---|---|---|---|
| 1 | Oslo og Akershus | 03 | Oslo | 69,307 | 1 |
|   |   | 02 | Akershus | 47,308 | 22 |
| 2 | Hedmark og Oppland | 04 | Hedmark | 22,122 | 26 |
|   |   | 05 | Oppland | 20,335 | 22 |
| 3 | Sør-Østlandet | 01 | Østfold | 25,043 | 18 |
|   |   | 06 | Buskerud | 27,012 | 21 |
|   |   | 07 | Vestfold | 22,410 | 14 |
|   |   | 08 | Telemark | 16,442 | 18 |
| 4 | Agder og Rogaland | 09 | Aust-Agder | 10,297 | 15 |
|   |   | 10 | Vest-Agder | 16,798 | 15 |
|   |   | 11 | Rogaland | 38,358 | 26 |
| 5 | Vestlandet | 12 | Hordaland | 41,128 | 33 |
|   |   | 14 | Sogn og Fjordane | 13,586 | 26 |
|   |   | 15 | Møre og Romsdal | 24,848 | 36 |
| 6 | Trøndelag | 16 | Sør-Trøndelag | 27,210 | 25 |
|   |   | 17 | Nord-Trøndelag | 14,750 | 24 |
| 7 | Nord-Norge | 18 | Nordland | 22,593 | 44 |
|   |   | 19 | Troms | 14,552 | 25 |
|   |   | 20 | Finnmark | 7,719 | 19 |
| 7 | NORWAY | 19 |   | 481,813 | 430 |

*Source:* Statistics Norway [57]

Table 1 lists the regions and counties. The Norwegian data is more finely grained than in the other studies. There are 430 units at the lowest (NUTS5) level of municipalities, whereas the Hungarian data had 168 sub-regions, the Dutch consisted of 90 postcodes, and the German had a total of 438 NUTS3 regions. Technology is indicated in our data using the two-digit sector classification of the (NACE[2]) which is also used by Statistics Norway [58, 59]. The organizational dimension will be indicated by company size in terms of the number of employees. Size of a company can be considered as a proxy of innovative dynamics (e.g., Pugh *et al.*, [60, 61] and Blau & Schoenherr, [62]). For example, small and medium-sized companies can be expected to operate differently from large-size multinational corporations. The data are divided into eight classes which are detailed in Table 2.

Table 2: Distribution of employees in the Norwegian data and corresponding uncertainties.

| Size | Number of employees | Number of companies | Probability | Uncertainty |
|---|---|---|---|---|
| 1 | 0 | 292,629 | 0.607 | 0.437 |
| 2 | 1-4 | 100,356 | 0.208 | 0.471 |
| 3 | 5-9 | 38,702 | 0.080 | 0.292 |
| 4 | 10-19 | 25,777 | 0.053 | 0.226 |
| 5 | 20-49 | 16,450 | 0.034 | 0.166 |
| 6 | 50-99 | 4,921 | 0.010 | 0.068 |
| 7 | 100-249 | 2,318 | 0.005 | 0.037 |
| 8 | >250 | 666 | 0.001 | 0.013 |
|   |   |   |   | 1.711 |

*Source:* Statistics Norway [57]

This table also provides the total number of companies in each of the classes at the national level. The probability distribution of the classes and the expected information contents of these distributions (see section 3.3) are given in the last two columns of the table. Companies without employees account for over 60.7% of the companies in Norway, in contrast to the Hungarian (29.8%), and Dutch data (19.7%). (The German study included neither this class of companies nor the number of self-employed in firms.)

---

[2]*Nomenclature générale des Activites économiques dans les Communautés Européennes*

*3.2 Knowledge intensity and high tech*

We follow the OECD classification for the various NACE codes into groups representing high-tech manufacturing (HTM), medium-tech manufacturing (MTM), knowledge-intensive services (KIS) and high-tech services (HTS) (Table 3).[59, p. 7; 19, p. 186].

Table 3: Classification of high-tech and knowledge-intensive sector according to Eurostat.

| *High-tech Manufacturing* | *Knowledge-intensive Sectors (KIS)* |
|---|---|
| **30** Manufacturing of office machinery and computers <br> **32** Manufacturing of radio, television and communication equipment and apparatus <br> **33** Manufacturing of medical precision and optical instruments, watches and clocks <br><br> *Medium-high-tech Manufacturing* <br><br> **24** Manufacture of chemicals and chemical products <br> **29** Manufacture of machinery and equipment n.e.c. <br> **31** Manufacture of electrical machinery and apparatus n.e.c. <br> **34** Manufacture of motor vehicles, trailers and semi-trailers <br> **35** Manufacture of other transport equipment | **61** Water transport <br> **62** Air transport <br> **64** Post and telecommunications <br> **65** Financial intermediation, except insurance and pension funding <br> **66** Insurance and pension funding, except compulsory social security <br> **67** Activities auxiliary to financial intermediation <br> **70** Real estate activities <br> **71** Renting of machinery and equipment without operator and of personal and household goods <br> **72** Computer and related activities <br> **73** Research and development <br> **74** Other business activities <br> **80** Education <br> **85** Health and social work <br> **92** Recreational, cultural and sporting activities <br><br> Of these sectors, **64, 72** and **73** are considered *high-tech services*. |

*Source:* Laafia [59, p. 7]; Leydesdorff *et al.* [19, p. 186].

A total of 43.5% of the Norwegian companies are in these knowledge-intensive sectors. This is well below the Dutch data (with 51.3%), but above the German data (33.2%). The ratios between high- and medium-tech manufacturing are 0.17 for Norway, 0.35 for the Netherlands and 0.61 for Germany. As noted above, we mainly focus on the geographical dimension and leave the decomposition in industrial sectors to a later study. However, the aforementioned information indicates a low level of high-tech manufacturing in Norway compared to the other nations which have been studied.

Figure 2: Fractions of various high- and medium tech companies in Norwegian counties (2008 data).

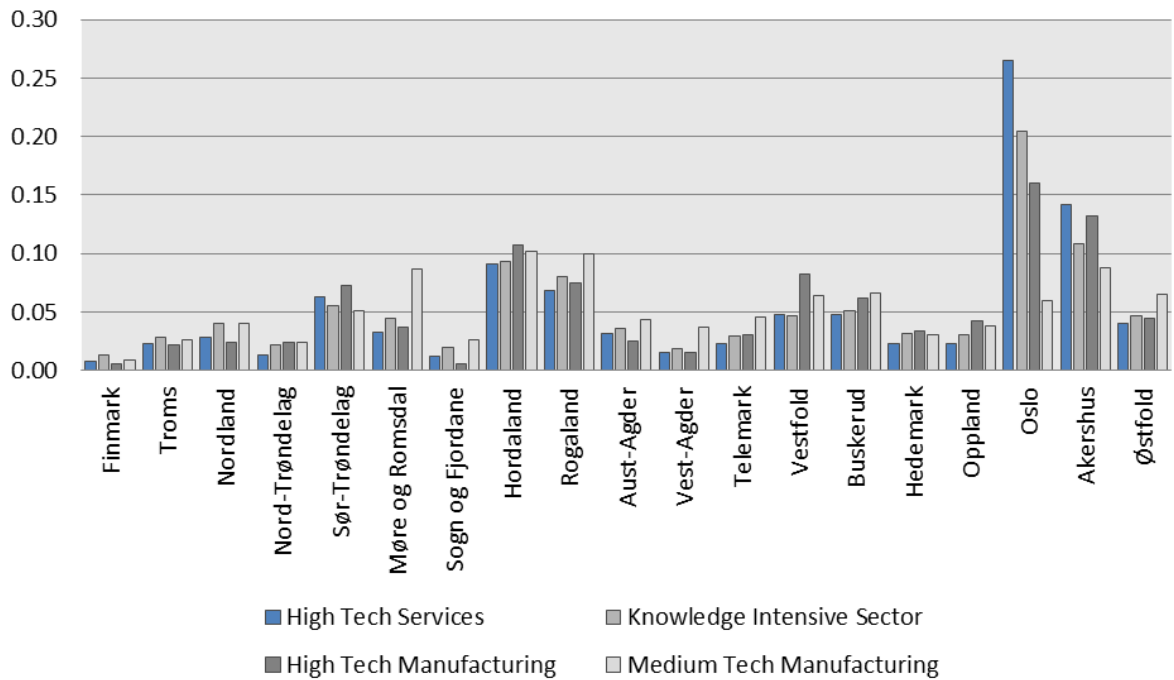

The fractions of various high- and medium-tech companies in the Norwegian counties are provided in Figure 2. As can be seen from this figure, HTS, HTM, and KIS are dominated by Oslo and Akershus. HTM is dominant in Vestfold, whereas MTM is dominant in Møre og

Romsdal and Rogaland. This indicates that the microsystem cluster in Vestfold and the maritime cluster in Møre og Romsdal can be identified [7]. Hordaland seems to score almost equally on all four fractions. The system engineering cluster in Buskerud can be identified as a high level of HTM and MTM. A more detailed analysis of the knowledge-intensive industries in Norway can be found in [56].

*3.3 Methodology*

According to Shannon [44] the uncertainty in the relative frequency distribution of a variable $x$ (that is, $\sum_X p_X$) is defined as $H_X = -\sum_X p_X \log_2 p_X$. Shannon denotes this as probabilistic entropy. Different from thermodynamic entropy, probabilistic entropy is dimensionless and therefore yet to be provided with meaning when a system of reference is specified. If one uses base two for the logarithm, then all values are expressed in bits of information.

Likewise, the uncertainty in a two-dimensional probability distribution can be defined as $H_{XY} = -\sum_X \sum_Y p_{XY} \log_2 p_{XY}$. In the case of interaction between the two dimensions, the uncertainty is reduced with the mutual information or transmission: $T_{XY} = (H_X + H_Y) - H_{XY}$. If the distributions are completely independent then $T_{XY} = 0$ and $H_{XY} = H_X + H_Y$. In the case of three interacting dimensions, the mutual information can be defined as follows [63]:

$$T_{XYZ} = H_X + H_Y + H_Z - H_{XY} - H_{XZ} - H_{YZ} + H_{XYZ} \qquad (1)$$

Krippendorff [64] noted that $T_{XYZ}$ can no longer be considered as Shannon-type information, since transmission, by definition, is linear and positive. It should be noted that the bilateral relations between the variables reduce the uncertainty, but that the trilateral term feeds back on this reduction and adds another term to the uncertainty. A negative uncertainty or information can also be considered as a redundancy.

Krippendorff [64] showed that the mutual information in three (or more) dimensions can be considered as the difference between redundancy and uncertainty generation at the network level among three (or more) subdynamics. In the Triple-Helix argument, the redundancy is generated by an overlay of communications in which different meanings can be translated into one another among academic, industrial, and political perspectives. Thus, more options can be generated endogenously; the maximum information content is enlarged when more options are made available. However, the interactions in the networks generate at the same time and necessarily uncertainty.[3] The difference between redundancy generation and uncertainty generations can be positive or negative. Leydesdorff [34] further elaborated the operationalization and measurement of information and redundancy in such configurations.

The value of $T_{GTO}$ measures the interrelatedness of the three sources of variance in this study and the fit of the relations between and among them. Because it is a measure of reduction of uncertainty, a better fit is indicated by a more negative value. This overall reduction of the uncertainty can be considered as a result of the intensity and the productivity of an innovative division of labour in a broad sense [20]. This transmission coefficient is also equal to the K-way interaction information (KWII) used by [40] when the sign is changed.

---

[3] Probabilistic entropy is coupled to thermodynamic entropy by the Boltzmann constant: $S = k_B * H$. Since $k_B$ is a constant with dimensionality Joule/Kelvin and H is dimensionless, the Second Law is equally valid for $H$ and $S$.

Our calculations contain three single-parameter uncertainties: a geographical $H_G$, a technological $H_T$, and an organizational $H_O$. The three two-parameter uncertainties are: $H_{GT}$, $H_{GO}$, and $H_{TO}$. The three-parameter uncertainty is denoted $H_{GTO}$. Similarly, the calculations contain three two-parameter transmissions ($T_{GT}$, $T_{GO}$, $T_{TO}$) and one three-parameter transmission $T_{GTO}$. The numerical results, however, are abstract and yet meaningless; they need to be appreciated using substantive theories. As noted, we appreciate the values of the bilateral transmissions as indicators of the three knowledge functions specified above that may lead to synergy in one configuration more than in another. We enrich the discussion further with other concepts, but one should be aware that this appreciation only has the status of stimulating the heuristics by raising questions for further research suggested by our results,

*3.4 Statistical decomposition*

One of the advantages of information theory is that the values are based on summations and can therefore be fully decomposed. Analogous to the decomposition of Shannon-type information [25], the mutual information can be decomposed into groups as follows:

$$T = T_0 + \sum_i \frac{n_i}{N} T_i \tag{2}$$

Since we decompose in the geographical dimension, $T_0$ will be in-between county uncertainty, $T_i$ the uncertainty prevailing in each county $i$, $n_i$ is the number of firms in this county, and $N$ the total number of firms in the whole country. The in-between group uncertainty ($T_0$) can be considered as a measure of the dividedness among the counties. A negative value of $T_0$ indicates additional synergy at the higher level of national (or regional) agglomeration among the counties. In the Netherlands, for example, such a surplus was found at the national level; in Germany, a surplus could not be retrieved at the national level, but it could be found at the level of the federal states (*Länder*). Note that one cannot compare the quantitative values of $T_0$ across countries—because these values are sample-specific—but one is allowed to compare the 'dividedness' in terms of the positive or negative signs of $T_0$.

*3.5 R&D expenditure*

Benner and Sandström [65] argued that institutionalization of a Triple Helix is critically dependent upon the form of research funding. The distribution of R&D expenditures over the 19 counties is listed in Table 4. The per capita R&D expenditure ranges from US$4600 [4] in Sør-Trøndelag to less than US$200 in Hedmark [66]. The budget distribution is very concentrated: the shares among counties range from 29.6% in Oslo and 15.8% in Sør-Trøndelag to below 1% for other counties. The industrial part of R&D expenditure is low in the main university counties (Oslo, Hordaland, Sør-Trøndelag, Akershus and Troms), but the lowest level is found in Finmark. The highest levels of industrial R&D funding can be found in Buskerud, Vestfold, and Telemark.

---

[4] Norwegian Kroner; NOK 1 is approximately equivalent to US$ 0.19.

Given this uneven geographical distribution of R&D funding, one would expect considerable spillover into science-based sectors from academic research institutions in the regions Oslo, Akershus, and Sør-Trøndelag. These regions absorb more than 55% of the total R&D funding in Norway. The 'institutional thickness' [67], as well as the high intensity of human capital in these regions [68], can be expected to generate favourable conditions for knowledge-based innovations.

Table 4: The total R&D expenditures in Norway (2007, 2009).

| County (NUTS3) | Industrial part of total R&D expenditures in each county[5] | Counties' share of total R&D expenditures in Norway[6] | R&D expenditure per capita by county (NOK) |
|---|---|---|---|
| Oslo | 37.5 % | 29.6 % | 22,411 |
| Østfold | 54.8 % | 1.9 % | 2,404 |
| Akershus | 57.1 % | 12.7 % | 11,255 |
| Hedmark | 41.5 % | 0.5 % | 940 |
| Oppland | 71.8 % | 1.5 % | 2,526 |
| Buskerud | 95.3 % | 4.8 % | 4,799 |
| Vestfold | 84.0 % | 2.3 % | 4,553 |
| Telemark | 77.8 % | 1.6 % | 4,639 |
| Aust-Agder | 58.8 % | 0.6 % | 4,177 |
| Vest-Agder | 69.7 % | 2.0 % | 4,177 |
| Rogaland | 71.1 % | 5.5 % | 4,799 |
| Hordaland | 29.7 % | 11.9 % | 9,855 |
| Sogn og Fjordane | 73.9 % | 0.8 % | 2,599 |
| Møre og Romsdal | 73.5 % | 2.1 % | 3,503 |
| Sør-Trøndelag | 33.1 % | 15.8 % | 24,094 |
| Nord-Trøndelag | 56.1 % | 0.8 % | 1,875 |

[5] 2007 data, NIFU-STEP [66]
[6] 2009 data, [66]

| | | | |
|---|---|---|---|
| Nordland | 55.4 % | 1.1 % | 2,057 |
| Troms | 11.1 % | 4.1 % | 12,187 |
| Finnmark | 7.8 % | 0.2 % | 1,498 |
| Total | 46.5 % | 100 % | |

Isaksen and Onsager [56] showed that the rates of firm creation are higher in these urban areas, but the firms are generally less innovative than in other parts of the country. They indicate that among the reasons for firms in small-urban and rural regions being more innovative than firms in urban areas is the much higher rate of public funding for innovative activities.

*3.6 The foreign factor*

Norway is a small and open economy, and foreign factors [22, 23] may play an important role in the Triple-Helix dynamics of its economy. Data on foreign direct investment (FDI) in Norway show that 36% of FDI in 2009 [69] is directed towards the oil and gas sector (NACE code 11), 18% towards real estate activities (NACE 70), and 13% towards the production of chemicals (NACE 24). The centre of the oil and gas industry is located in Rogaland and Hordaland. The investments in real estate are located in the large cities, whereas one expects investments in the chemical industry to be more geographically distributed across counties. FDI is expected to enhance technology transfer in the industry segment, but not in the real estate segment. Data on export value from Norwegian counties in 2008 show that Hordaland, Rogaland, and Møre og Romsdal have the highest export incomes [70].

4. **Results**

As noted above, the data can be (dis)aggregated in terms of geographical regions (NUTS2) and counties (NUTS3). The numbers of firms and municipalities in each county was provided above in Table 1. The number of firms is highest in Oslo with 69,307 firms and lowest in Finmark with 7,719 firms. Oslo contains only one unit at the NUTS5 level. This leads to $H_G = 0$ since there is no geographical uncertainty left. Consequently, $H_O$ would be equal to $H_{GO}$ and $H_{TO}$ to $H_{GTO}$ and no synergy can then be calculated for Oslo. This problem, however, can be overcome by the calculations at a higher level of aggregation (NUTS2), in which case the data from Oslo and Akershus are combined; the synergy is consequentially unequal to zero.

*4.1 Uncertainty at the county level*

Table 5 shows the uncertainty in the geographical distribution at the NUTS3 level in the first column. This indicator of the geographical concentration of economic activities has the highest value for Nordland, in this case 4.783 bits, which equals to 87,6% of the maximum entropy for a county with 44 municipalities ($\log_2(44)=5.46$). In other words, the economic activity is most decentralized in this county. If we use this percentage of the maximum information content as a decentralization parameter, the highest value—94.9% of the maximum information content —is found in Sogn og Fjordane with its 26 municipalities.

The most centralized counties are Hordaland with 63.3% of maximum uncertainty and Sør-Trøndelag with 66.0%. Both counties are characterized by one large city—Bergen in Hordaland and Trondheim in Sør-Trøndelag—and a number of small surrounding municipalities. In these two counties, more than 50% of the population is located in the large city. In order to be able to compare the various counties with different numbers of municipalities, we scale the information values as a percentage of the maximum uncertainty. This is relevant for all parameters which include geographical parameters.

Table 5: Information contents (in bits) of the distributions in three dimensions and their combinations at NUTS3 level.

| Name | $H_G$ | $H_T$ | $H_O$ | $H_{GT}$ | $H_{GO}$ | $H_{TO}$ | $H_{GTO}$ |
|---|---|---|---|---|---|---|---|
| Finmark | 3.771 | 4.337 | 1.846 | 7.810 | 5.583 | 5.902 | 9.177 |
| Troms | 3.427 | 4.337 | 1.819 | 7.559 | 5.230 | 5.938 | 9.008 |
| Nordland | 4.783 | 4.336 | 1.777 | 8.865 | 6.531 | 5.862 | 10.204 |
| Nord-Trøndelag | 3.970 | 3.985 | 1.619 | 7.798 | 5.569 | 5.329 | 8.993 |
| Sør-Trøndelag | 3.066 | 4.270 | 1.736 | 7.069 | 4.783 | 5.809 | 8.498 |
| Møre og Romsdal | 4.678 | 4.357 | 1.776 | 8.784 | 6.432 | 5.888 | 10.144 |
| Sogn og Fjordane | 4.462 | 4.026 | 1.632 | 8.285 | 6.065 | 5.414 | 9.481 |
| Hordaland | 3.192 | 4.301 | 1.752 | 7.283 | 4.933 | 5.876 | 8.755 |
| Rogaland | 3.792 | 4.226 | 1.757 | 7.800 | 5.535 | 5.785 | 9.258 |
| Aust-Agder | 2.939 | 4.290 | 1.717 | 7.052 | 4.648 | 5.791 | 8.455 |
| Vest-Agder | 3.205 | 4.412 | 1.741 | 7.466 | 4.931 | 5.930 | 8.849 |
| Telemark | 3.588 | 4.350 | 1.729 | 7.781 | 5.301 | 5.856 | 9.163 |
| Vestfold | 3.276 | 4.240 | 1.713 | 7.404 | 4.976 | 5.780 | 8.859 |
| Buskerud | 3.855 | 4.240 | 1.676 | 7.942 | 5.518 | 5.745 | 9.340 |
| Hedemark | 4.415 | 3.954 | 1.567 | 8.221 | 5.964 | 5.289 | 9.428 |
| Oppland | 4.049 | 4.092 | 1.625 | 7.993 | 5.653 | 5.505 | 9.284 |
| Oslo | 0.000 | 4.025 | 1.669 | 4.025 | 1.669 | 5.558 | 5.558 |
| Akershus | 3.942 | 4.187 | 1.689 | 7.998 | 5.619 | 5.732 | 9.466 |
| Østfold | 3.517 | 4.237 | 1.714 | 7.614 | 5.215 | 5.768 | 9.057 |
| Norway | 7.275 | 4.319 | 1.711 | 11.317 | 8.960 | 5.856 | 12.729 |

The maximum information content of the technological distribution ($H_T$) is $\log_2(60)=5.91$, and $\log_2(8)=3$ in the organizational dimension ($H_O$). The highest level for the uncertainty in the technology distribution is found in Vest-Agder with 74.7% of maximum information content and the lowest in Hedmark with 66.7%. This indicates that the industry variation is lowest in Hedmark where primary industries dominate.

The uncertainties for all counties vary moderately in the Norwegian data, but are higher than for the Netherlands [19] (after normalization as percentages of the maximum uncertainty). This indicates that the industry variation in Norway is larger than in the Netherlands. The uncertainty of the organizational distribution is largest in Finmark with 61.5% of the maximum information content and lowest in Hedmark with a value of 52.2%. Both counties are characterized by strong primary industries; however, in Hedmark this is due to agriculture, whereas in to Finmark fishing and fish-processing are expected to dominate. The agricultural sector is characterized by a large number of small units where farmers often organize their farms into several companies. The values for these counties ranging from 52.2%-61.5% indicate a highly skewed distribution. This can be seen in Table 2: the distribution is dominated by small companies.

The combined uncertainties in two dimensions ($H_{GT}$, $H_{GO}$, $H_{TO}$) reduce the uncertainty at the systems level (Equation 1). $H_{GT}$ is highest in Møre og Romsdal and lowest in Sør-Trøndelag with 4.78 bits. This suggests that there is a weaker link (and thus more interaction across the distribution) between geography and technology in Møre og Romsdal (more diversified economy) than in Sør-Trøndelag. In this latter region, most technological firms are expected to be located in Trondheim, closely linked to the Technical University.

$H_{TO}$ is highest in Troms and Vest-Agder, which has Norway's highest level of combinations of technological and organizational specialization. The lowest value for this indicator is found in Hedmark. This is probably due to the primary industries in this county. $H_{GO}$ has the highest value in Sogn og Fjordane and in Møre og Romsdal, indicating that firms of all sizes are distributed across these counties. The lowest value is found in Oslo, but this is caused by the lack of uncertainty in the geographical distribution.

Table 6: The mutual information contents (in mbits) of the distributions in three dimensions at NUTS3 level.

| Name | $T_{GT}$ | $T_{GO}$ | $T_{TO}$ | $T_{GTO}$ | $\Delta T_{GTO}$ in mbits |
|---|---|---|---|---|---|
| Finmark | 0.298 | 0.035 | 0.281 | -0.163 | -2.617 |
| Troms | 0.206 | 0.016 | 0.218 | -0.135 | -4.076 |
| Nordland | 0.254 | 0.029 | 0.251 | -0.158 | -7.392 |
| Nord-Trøndelag | 0.158 | 0.021 | 0.275 | -0.128 | -3.924 |
| Sør-Trøndelag | 0.267 | 0.019 | 0.197 | -0.092 | -5.175 |
| Møre og Romsdal | 0.251 | 0.022 | 0.245 | -0.149 | -7.702 |
| Sogn og Fjordane | 0.202 | 0.029 | 0.245 | -0.162 | -4.579 |
| Hordaland | 0.210 | 0.010 | 0.176 | -0.093 | -7.973 |
| Rogaland | 0.218 | 0.014 | 0.198 | -0.087 | -6.947 |
| Aust-Agder | 0.177 | 0.009 | 0.216 | -0.089 | -3.106 |
| Vest-Agder | 0.151 | 0.015 | 0.223 | -0.121 | -2.576 |
| Telemark | 0.157 | 0.016 | 0.222 | -0.109 | -3.701 |
| Vestfold | 0.111 | 0.013 | 0.173 | -0.073 | -3.396 |
| Buskerud | 0.153 | 0.012 | 0.171 | -0.095 | -5.304 |
| Hedemark | 0.147 | 0.019 | 0.232 | -0.111 | -5.093 |
| Oppland | 0.148 | 0.022 | 0.211 | -0.101 | -4.282 |
| Oslo | 0.000 | 0.000 | 0.136 | 0.000 | 0.000 |
| Akershus | 0.131 | 0.012 | 0.144 | -0.064 | -6.330 |
| Østfold | 0.140 | 0.016 | 0.183 | -0.072 | -3.730 |
| Norway | 0.277 | 0.025 | 0.174 | -0.100 | -99.594 |
| | | | | Sum | -87.919 |
| | | | | $T_0$ | -11.675 |

The data on the various transmission coefficients for the counties are provided in Table 6. Using Equation 1, the synergy is scaled—in the right-most column—with the number of firms in each county in order to find their contribution to the national level. We scaled the standardized synergy from bits to millibits (mbits) in order to enhance the readability.

Table 6 shows that 11.7% of the uncertainty at the national level is generated between the counties. Furthermore, there is more mutual information between the geographical distribution of firms in Norway and their technological specialization than between the geographical distribution and their size ($T_{GT}$=0.277 bits, compared to $T_{GO}$=0.025 bits). The mutual information between technology and organization is larger than $T_{GO}$, but smaller than $T_{GT}$. $T_{GO}$ and $T_{GT}$ have also been considered as indicators of geographical clustering [19].

The lowest $T_{GT}$ values are found in the counties surrounding Oslo, indicating a diversified industry structure, as may be expected in the neighbourhood of the largest city and capital. The highest values for this parameter, as well as for $T_{GO}$, occur in the northernmost counties. This indicates more specialized industry. Nordland shows a value that is an order of magnitude higher than for the rest of the counties. This must be due to the specific geography and the large number of small municipalities in this county.

The $T_{TO}$ parameter can perhaps be appreciated as a correlation between the maturity of the industry and the size of the firms involved. The lowest values for this parameter occur in the metropolitan area in the counties Oslo and Akershus. These values indicate a less mature techno-economic structure in these counties. The highest values of this parameter occur in the northern counties. These high values may indicate an over-mature techno-economical structure. The number of small companies is low in these areas. The dynamics of the companies in these counties may have been altered due to regional economic measures such as various subsidies and tax reliefs. The many small municipalities in the northern part of Norway require a relatively large public sector. The percentage of the population occupied in the public sector in 2008 is 40% in the northern counties as compared with 30% in rest of the country [71].

Figure 3: Contributions to the knowledge base of the Norwegian economy of the 19 counties at the NUTS3 level.

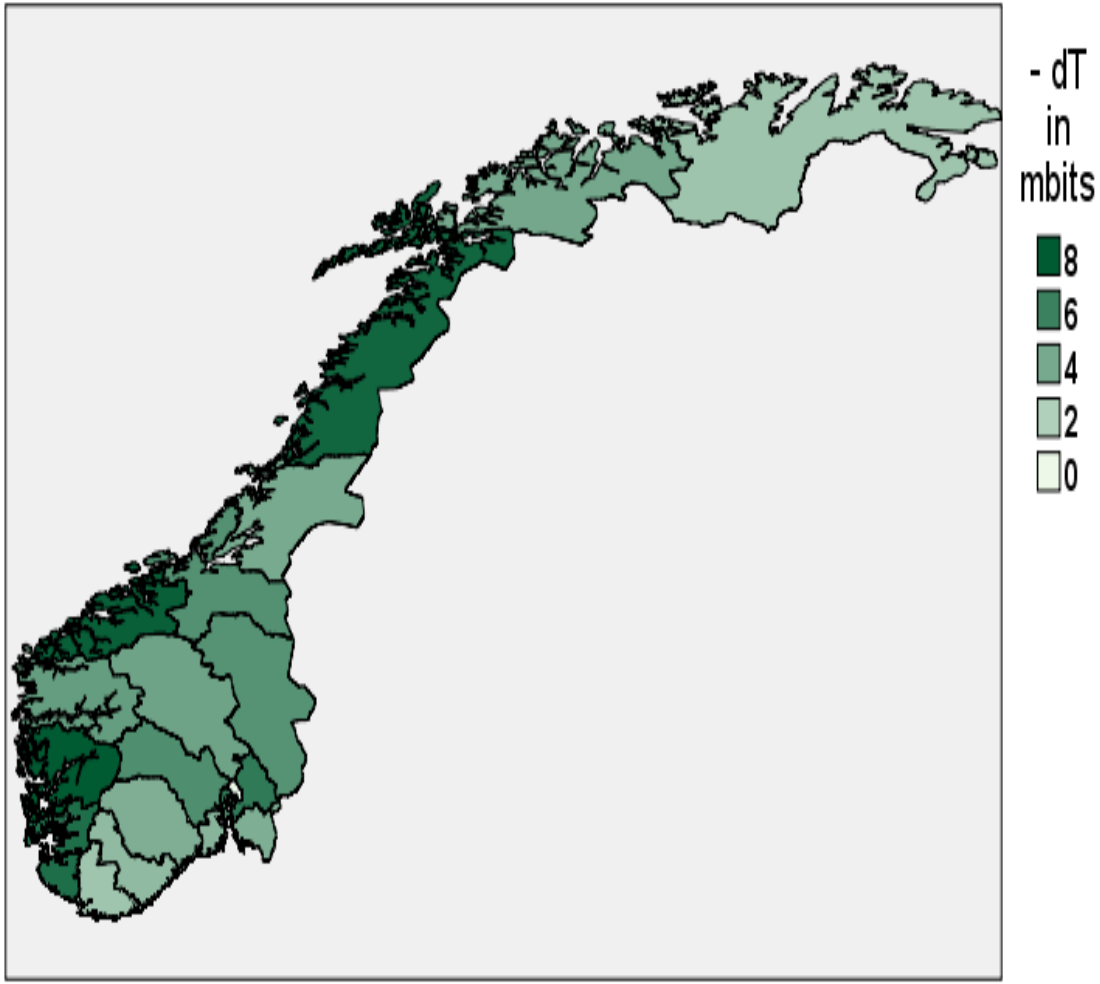

The synergy among the three knowledge functions ($T_{TGO}$) is highest in Hordaland (-7.97 mbits), Møre og Romsdal (-7.70 mbits), Nordland (-7.41 mbits) and Rogaland (-6.95 mbits). These results are shown in Figure 3. These counties are characterized by a strong industry sector, dominated by oil and gas production, and maritime industries.

*4.2 Uncertainties at the regional level*

In order to analyze the effects of including the capital Oslo, we repeated the analysis at the next-higher level of seven NUTS2 regions. Are the trends similar? The composition of the counties in each region was given in Table 1 above, and the results are provided in Table 7. The in-between contribution when aggregating from the regions to the national level is 2.7% of the total synergy. In other words, this percentage of the synergy is to be found above the regional-level.

Table 7: The information and mutual information contents (in mbits) of the distributions in three dimensions at NUTS2 level.

|  | Norway | Oslo og Akershus | Hedemark og Oppland | Sør-Østlandet | Agder og Rogaland | Vestlandet | Trøndelag | Nord-Norge |
|---|---|---|---|---|---|---|---|---|
| N | 481,819 | 116,610 | 42,457 | 90,908 | 65,453 | 79,562 | 41,961 | 44,864 |
| $H_{GTO}$ | 12.729 | 8.118 | 10.357 | 11.088 | 10.363 | 10.765 | 9.607 | 11.102 |
| $H_G$ | 7.275 | 2.573 | 5.239 | 5.547 | 4.856 | 5.325 | 4.319 | 5.632 |
| $H_T$ | 4.319 | 4.143 | 4.026 | 4.273 | 4.304 | 4.314 | 4.203 | 4.348 |
| $H_O$ | 1.711 | 1.678 | 1.595 | 1.706 | 1.745 | 1.740 | 1.696 | 1.804 |
| $H_{GT}$ | 11.317 | 6.611 | 9.111 | 9.667 | 8.931 | 9.375 | 8.261 | 9.722 |
| $H_{GO}$ | 8.960 | 4.246 | 6.813 | 7.239 | 6.587 | 7.046 | 5.995 | 7.408 |
| $H_{TO}$ | 5.856 | 5.688 | 5.405 | 5.805 | 5.853 | 5.856 | 5.684 | 5.918 |
| $T_{GT}$ | 0.277 | 0.105 | 0.154 | 0.154 | 0.229 | 0.264 | 0.262 | 0.258 |
| $T_{GO}$ | 0.025 | 0.006 | 0.020 | 0.015 | 0.014 | 0.019 | 0.021 | 0.028 |
| $T_{TO}$ | 0.174 | 0.133 | 0.216 | 0.174 | 0.196 | 0.198 | 0.216 | 0.235 |
| $T_{GTO}$ | -0.100 | -0.033 | -0.112 | -0.096 | -0.103 | -0.134 | -0.113 | -0.162 |
| mbits | -99.594 | -7.884 | -9.858 | -18.058 | -14.049 | -22.108 | -9.847 | -15.104 |
| Sum | -96.910 | | | | | | | |
| $T_0$ | -2.687 | | | | | | | |

Not surprisingly, the most centralized region is Oslo og Akershus with a $H_G$ of 56.9% of the maximum information content and the most decentralized is Hedmark og Oppland with a $H_G$ of 93.8%. The economic activity in the inland region Hedmark og Oppland is more decentralized than the northern region (Nord-Norge) due to the fact that the municipalities are more equal in size.

The uncertainty in the technological distribution ($H_T$) ranges from a lowest value for Hedmark og Oppland to the highest values which occur in Nord-Norge and Vestlandet. This indicates that the industry structures are slightly more diversified in the latter regions. The uncertainty in the organizational distribution is highest in Nord-Norge and lowest in Hedemark og Oppland. The large number of small business units in an agriculture-dominated region like Hedmark og Oppland is thus contrasted with the relatively larger number of medium-sized units in the fish and fish-farming dominated region of Nord-Norge.

With regard to the knowledge functions, the knowledge exploration ($H_{GT}$ parameter) is highest in Sør-Østlandet and lowest in Oslo og Akershus. This indicates a more diversified industry structure where companies in most industries are found all over the region. The organizational control ($H_{GO}$ parameter) is highest in Hedmark og Oppland and lowest in Oslo og Akershus. This indicates that companies of all sizes are distributed all over Hedmark og Oppland, whereas in Oslo og Akershus the size and the geographical distribution are better correlated.

The contribution to the synergy across knowledge functions at the regional level is shown in Figure 4. It is highest for Vestlandet (-22.1 mbits) and lowest for Oslo og Akershus (-7.8 mbits). The inter-regional contribution to the national level is only 2.7% of the total synergy. This indicates that the main contribution to the synergy comes from the aggregation at the regional, rather than the national level. Somewhat unexpectedly, the synergy is low in Trøndelag and Oslo og Akershus, where the main knowledge institutions are located, and high in the industrial regions at the west coast. The high synergy in Nord-Norge was an unexpected result.

Figure 4: Contribution to the knowledge base of the Norwegian economy from regions at NUTS2 level.

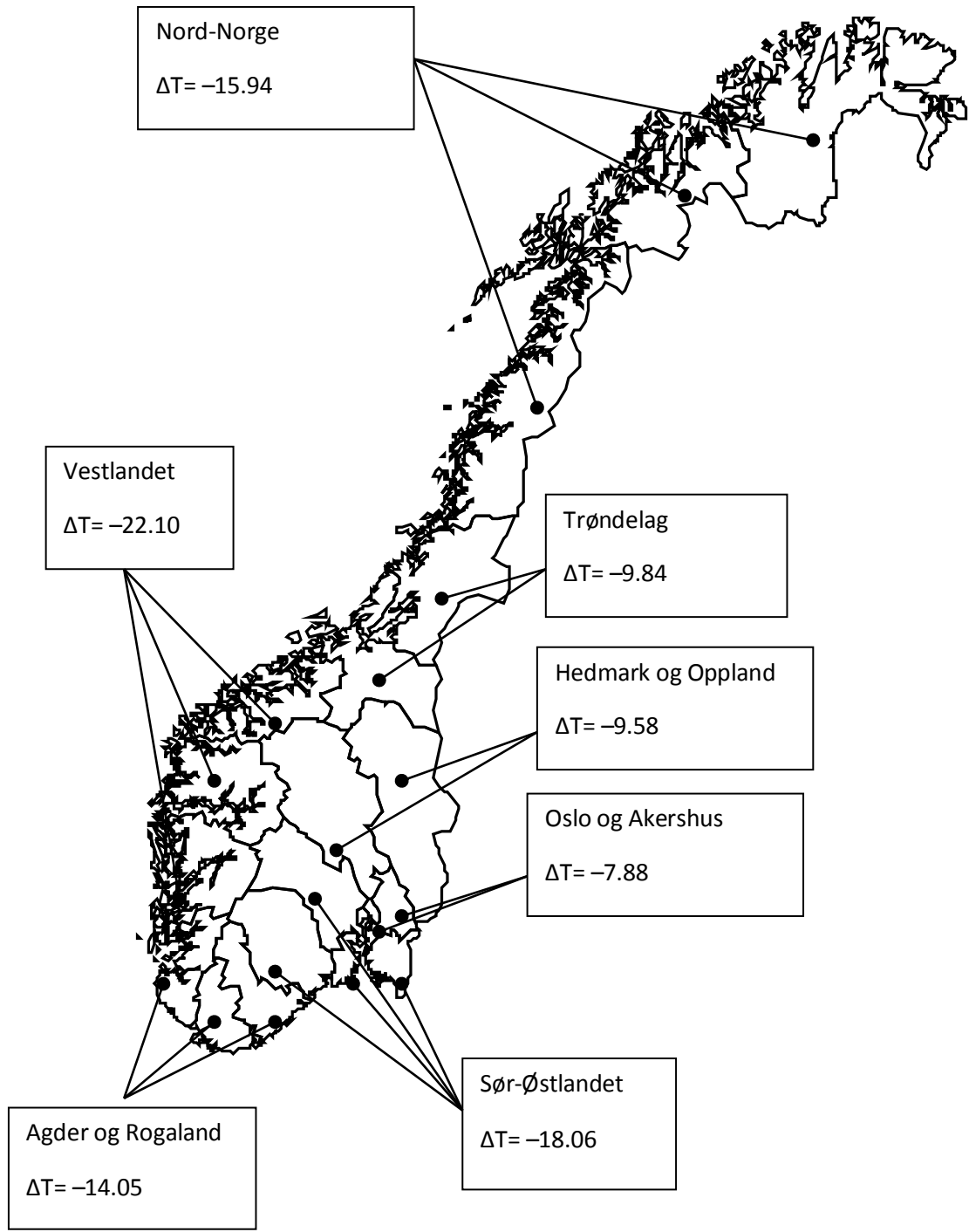

The industry in the counties on the west coast is internationally oriented. This raises questions as to whether international knowledge spill-overs from customers are perhaps more important than local knowledge spill-over from academic institutions [72]. A high export value from a county indicates an internationally oriented industry (section 3.6). The largest industrial companies in Rogaland are dominated by companies in the oil and gas sector. The domestically owned multinational company Statoil, and national branches of foreign owned multinational companies like Total, Esso, and Exxon Mobil, are the largest companies in this county. The two largest industrial companies in Møre og Romsdal—Rolls-Royce Marine and Stx Osv—have foreign owners.

Maritime offshore projects are often characterized by a high degree of customization with extended cooperation between customers and suppliers, since both sides are part of global value chains. This leads to knowledge spillover from global sources of knowledge to the various participants in these projects. There is also a significant and substatntial amount of customised training and job rotation, adding to the diffusion of knowledge.

## 5. Conclusions and discussion

When analysing the Norwegian economy in terms of Triple-Helix synergies, we find a similar pattern at different geographical scales. These results suggest that the counties and regions that contribute most to the knowledge base of the Norwegian economy are located on the western coast of Norway. Within the framework of the Triple-Helix theory, these areas seem to have achieved a balance between the three sub-dynamics to a larger extent than other parts of the country. In the northern part of the country government intervention is so substantial that the dynamics of the economy are changed. This can best be seen by the lack of new small companies and the high level of public employees (40%) in these counties. The exception is Tromsø, the main university city in the north, where the number of start-ups is high. One of the reasons may be the government's focus on marine biotechnological research at this university. However, most of the marine industry is located in Vestlandet.

Most of the research capacity in Norway is located in Oslo and Trondheim, in areas with weak industrial traditions. The industrial counties on the west coast are characterized by a strong internationally oriented manufacturing industry directed towards maritime, offshore and marine industry. These firms operate in global markets. The knowledge base is synthetic [7], with a low share of formal higher education. Møre og Romsdal contains the strongest industry cluster in Norway: the maritime cluster. The high-tech clusters, located in other parts of the country are probably too small to influence the synergy at the NUTS3 level significantly. At the NUTS2 level, the highest level of synergy is also found in Vestlandet. This shows that our results are robust against changes in the geographical scale.

There are some interesting differences between the geographical influence on the results in the case of Norway or the Netherlands. Whereas in the data from the Netherlands [19], the geographical uncertainty is correlated with the number of firms in the region ($r = 0.76$), the Norwegian geographical uncertainty correlates negatively with the number of firms ($r = -0.61$). The comparison between public R&D expenditure and the synergy of the knowledge base provides another negative correlation in the case of Norway. In our opinion, these findings confirm the conclusions of Onsager *et al.* [28] and OECD [32] that areas in Norway with high concentrations of knowledge institutions (and hence a high level of higher education) seem to live in 'separate' worlds, uncoupled from the needs of the industry. Easy access to public research funding through networks and co-location with research councils and political decision makers makes the transaction costs of engagement with fellow academics lower than those with industry [31].

At the national level, Shelton and Leydesdorff [73] found that high levels of private R&D funding promote cooperation with industry and results in a larger numbers of patents. A high level of public funded R&D results in an increased number of academic papers. This underpins the findings of Benner and Sandström [65] that institutionalization of a Triple-Helix model is critically dependent upon the form of research funding. There is also a tendency in the academic literature to fail to see the importance of innovation in 'low-tech' industries [74].

Foreign factors, such as high FDI, foreign ownership and global customers, are characteristic for the regions and counties with the highest synergy. This may support what Bathelt *et al.* [75] called a 'local buzz - global pipeline' effect, that is, a combination of geographically embedded local knowledge with knowledge from global sources, filtered for relevance by global customers. The dominating industry sectors in these littoral counties are medium-tech manufacturing. Easy access to local tacit knowledge and international knowledge spillovers from customers may be more important than codified academic knowledge. Calculation of the inter-group synergy consequently indicates that synergy occurs at the regional, rather than at the national level.

Our results support the findings from previous studies showing that medium-tech manufacturing rather than high-tech manufacturing is associated with synergy [19, 20, 76]. Our results also show the effect of high levels of government intervention in the northern part of the country [1]. In these regions, our measures were dependent on the scale of the aggregation (NUTS2 or NUTS3). However, public R&D funding is directed towards academic institutions in university cities, whereas regional policies are mostly directed towards the northern region and regions with little industry. The highest synergy in the knowledge functions in the Triple-Helix dynamics is to be found in the industrial counties on the west coast, where medium-tech manufacturing is concentrated and foreign factors associated with operating in global markets enhance synergy to a greater extent than expected.